\documentclass[12pt,dvips]{article}

\usepackage{amsmath,amssymb,exscale}
\usepackage{array,multicol}
\usepackage{afterpage,float,flafter}
\usepackage{epsfig,rotating,pifont}
\usepackage{cite}
\@addtoreset{equation}{section} \makeatother

\def\beq{\begin{eqnarray}}
\def\eeq{\end{eqnarray}}

\def\lsim{\mathrel{\rlap{\lower3pt\hbox{\hskip0pt$\sim$}}
    \raise1pt\hbox{$<$}}}         
\def\gsim{\mathrel{\rlap{\lower4pt\hbox{\hskip1pt$\sim$}}
    \raise1pt\hbox{$>$}}}         

\title{
\vspace{1cm}
\Large\textbf{Black Holes and Large $N$ Species  Solution to the Hierarchy Problem}
\vspace*{.5cm}
\author{
\large \textbf{Gia Dvali\footnote{email: gd23@nyu.edu}}\\
\emph{Center for Cosmology and Particle Physics,}\\\emph{Department of Physics, New York University}\\
\emph{4 Washington Place, New York, NY 10003}}}

\date{}
\begin{document}
\maketitle \thispagestyle{empty} \vspace*{.5cm}

\begin{abstract}
We provide the perturbative and non-perturbative arguments showing that theories with 
large number of  species of the quantum fields,  imply
an inevitable hierarchy between the masses of the species  and the Planck scale, shedding a different light on 
the  hierarchy problem. In particular, using the black hole physics, we prove that  any consistent theory that includes $N$ $Z_2$-conserved species of the quantum fields of mass 
$\Lambda$,  must have a value of the Planck mass, which in large $N$ limit is  given by  $M_P^2 \gsim N\Lambda^2$. 
An useful byproduct of this  proof is that any exactly conserved quantum charge, not associated with a long-range classical  field, must be defined maximum modulo $N$, with $N\, \lsim \, (M_P/m)^2$, where 
$m$ is the mass of the unit charge. 
For example, a  continuous global  $U(1)$ `baryon number' symmetry, must be explicitly broken by gravity, at least down to a $Z_N$ subgroup,  with $N \lsim  (M_P/m_b)^2$, where 
$m_b$ is the baryon mass.  The same constraint applies to any discrete  gauge symmetry, as well as to other quantum-mechanically-detectable black hole charges that are associated  with the  massive quantum hair of the black hole. We show that  the gravitationally-coupled $N$-species sector that solves the gauge hirearchy problem, should be probed by LHC. 

\end{abstract}

\newpage
\renewcommand{\thepage}{\arabic{page}}
\setcounter{page}{1}

\section{Introduction}

 The essence of the hierarchy problem is an inexplicably large separation between the weak and the
 Planck ($M_P$)  scales, or equivalently between  $M_P$ and the Higgs mass.   The latter, is quadratically sensitive to the cut-off of the theory, and this is the source of the problem.  
Either the cut-off is very high (say $\sim M_P$), and then one is left with the question, what stabilizes the Higgs mass?  Or,  the cut-off is not far from the weak scale, but then one has to explain why gravity is so weak.  In both cases, the solution should come up in the form of an UV-insensitive large number, that 
sets the hierarchy between the scales.  For example, in supersymmetry, this number is the ratio of  
the the Plack mass to the supersymmetry breaking scale, whereas in Large Extra Dimensions
scenarion\cite{ADD}, the hierarchy is set  by the large volume of the extra space (in fundamental Planck units).  
 
  In the present paper, we shall discuss a different approach in which a large UV-stable number
  that provides the hierarchy between the Planck mass and a low energy scale $\Lambda$, is the 
  the number of the quantum field species, $N$. 

 We shall show that in the theories in which the number of species at the scale $\Lambda$ is $N \, \gg \, 1$, 
 the Plack mass must satisfy 
  \begin{equation}
\label{PM}
M_P^2 \, \gsim \, N \, \Lambda^2,
\end{equation}
 up to a factor that scales as $\sim$ ln$(N)$ with large $N$.
 We shall provide the two types of arguments supporting  this bound.  The first one is an already existing perturbative argument, which shows that 
 having $M_P^2 \ll N\Lambda^2$ for large $N$ is technically unnatural. 
 
  The second one is an exact  non-perturbative argument, which shows that  having a Planck mass 
 that violates the bound (\ref{PM}) is inconsistent with the black hole physics.  In the other words, we are lead to the conclusion that  the large number of species automatically weakens gravity by a  
 $1/N$-factor!
 
  Then, the following `cheap' solution to the hierarchy problem emerges.  Postulate the existence of 
 $N \sim 10^{32}$ quantum field species beyond the Standard Model.  For instance, $N$ copies of the 
 Standard Model related by the permutation symmetry.  Although, a low energy observer from  each Standard Model replica would be puzzled by the smallness of the weak scale versus the Planck mass, the hierarchy  would be guaranteed by the consistency of the theory with the black hole physics.

  
 The  proof of the bound (\ref{PM})  delivers an important byproduct, which relates the periodicity of  the conserved quantum numbers, to the allowed values of the Planck mass, in a model-independent way. 
 In particular, any conserved quantum number $Q$, which is not associated with a long-range 
 classically-observable  force, and is carried by the state(s) of mass $\Lambda$, can have a maximal
 periodicity  $N$ given by  (\ref{PM}). 
 
  For example,  this constraint automatically  applies to  the gauged discrete $Z_N$-symmetries,  as well as  to any exact subgroup of the continues global symmetries, in general. It also automatically results into the quantization of the black hole quantum charge associated with a  massive quantum hair.  Two types of the quantum hair are of our interest. 
 
   The fact that the black holes can carry a quantum mechanical hair under the discrete gauge symmetries, was discovered some time ago by Krauss and Wilczek \cite{ZN}. Such a quantum  hair
 results whenever a continuous gauge group is spontaneously broken (Higgsed) down to a discrete subgroup at an arbitrarily high scale.  Because of its $Z_N$-nature, such a  hair automatically falls within our constraint.  
 
    Interestingly,  the above bound (\ref{PM}) also constraints a different type of the black hole quantum hair.  In  \cite{quantum},  it was shown that the  black holes can carry a  quantum-mechanical hair under the massive gauge fields of an arbitrary integer spin, such as spin-2, or higher.  
   Naively, the corresponding charge is continuous and can take an arbitrary periodicity.  However, our arguments show that this is not true. 
   Since just like the $Z_N$-case, the massive spin-2 (or higher) hair is not supported by any classically-detectable long-range field,  the corresponding charge automatically falls within the validity range 
of our proof, and must be $N$-periodic, with $N$ satisfying the  bound (\ref{PM}). Thus,  the  quantum-mechanically detectable black hole charges always have limited periodicity bounded by (\ref{PM}).

 Thus, the number of species in the large $N$ solution of the hierarchy problem, can equally be replaced by a  large $N$-cycle  of a  $Z_N$ gauge symmetry,  or with  a large $N$-periodicity of the black hole charge under some  massive quantum hair.  

  As one of the central points,  we shall discuss a  phenomenologically-interesting observation,  that  a hidden sector with $N$-species, that  guaranties the hierarchy   between the Planck mass and the TeV scale , must be probed in the Standard Model particle collisions at energies  $\sim$ TeV.   

   Finally, the String Theory constructions often involve many species of quantum fields at low  
  energies.  This fact may be a blessing for stabilizing the various  scales in String Theory. 
 
  For example,  one can ask,  whether a large number of species could be a viable way of explaining the hierarchy  between the Planck mass and the Grand Unified  Theory (GUT) scales, since the realistic theories  often incorporate large number of states at  the GUT scale.

\section{Perturbative Renormalization of the Planck Mass} 

 We start with the perturbative argument first. This was suggested in \cite{DGKN} as a possible 
 many-species solution to the hierarchy problem in the context of the brane world gravity \cite{DGP}.  
The idea being that on a 3-brane populated by the $N$ four-dimensional species, the effective Planck mass  should scale  as $N$, even if the cutoff (which can be  either the Standard Model scale or the fundamental high-dimensional Planck mass) is fixed.    Our perturbative argument is exactly the same, although we wish to abstract from any particular high-dimensional realization and reduce the idea to its bare essentials in four-dimensions.    

 The point then is the following. Imagine  $N$ species of quantum fields, with masses at the scale $\Lambda$,  coupled to gravity.   Each of these species will contribute into the renormalization of the 
 Planck mass (equivalently,  the graviton wave-function renormalization)  the factor 
 $\sim \Lambda^2$\cite{adler}.  Neglecting the accidental cancellations,  this has to be multiplied by the number 
 of species, and  as a result the effective contribution to the Planck mass is $\sim N\Lambda^2$. 
In the other words, the perturbative renormalization arguments suggest that in the presence of 
$N$-species  gravity must be weakened by $1/N$.

\section{The Black Hole Proof} 

\subsection{The Leading Order Relation} 

  We shall now discuss a non-perturbative argument, based on the black hole physics, 
  that proves the bound (\ref{PM}).   Consider  $N$ species of the bosonic quantum fields $\Phi_j, ~~j= 1,2, ...N$,  of mass $\Lambda$. We shall first assume that the system is invariant under an {\it exact} 
  discrete  $Z_2^N\, \equiv\, Z_2^{(1)} \times Z_2^{(2)} \times \, ... \,Z_2^{(N)}$  symmetry, under the independent sign flips of the fields.  That is,  under any given
  $Z_2^{(j)}$-factor only one particular field changes the sign,  $\Phi_j \rightarrow  - \Phi_j$, 
  whereas all the other fields  are invariant. 
 
  We shall now prove that  in such a case the Planck mass must satisfy the bound (\ref{PM}).   
  
 

 In order to  prove the relation (\ref{PM}) we can perform the following thought experiment. 
 Taking an arbitrarily large number of  $N$-species  particles, we can prepare an arbitrarily large  black hole. 
 This black hole will carry the information about the amount of the conserved charge carried by 
 the particles.  In order to avoid  entering the discussions on the black hole information loss issues, 
 it is useful to think of these $Z_2$-s as the gauged discrete symmetries\cite{ZN}.  The information about the 
 absorbed charge then can be monitored by the Aharonov-Bohm effect at infinity, using the probe 
 $Z_2$-cosmic strings\cite{ABmassive},  and cannot be lost.
  
 Because the conserved quantum number is $Z_2^N$, we can store maximum 
 $N$ units of the charge in such a black hole. For this we will need $N$ particles, each belonging to a different species.  Any further increase of  the  number of the initial particles, will not increase the amount 
 of the conserved discrete charge stored by the black hole.  Thus, we shall focus on a minimal size black hole carrying the maximum possible discrete charge. 
 
  The mass of such a black hole 
 is 
 \begin{equation}
\label{bhmass}
M_{BH} \, = \, N\Lambda  
\end{equation} 
 
 Because of the conservation, the information about the $Z_2^N$-charge hosted by the black hole, must be revealed after its 
 evaporation.  For a black hole of the Hawking temperature $T_H$, the probability of the emission 
 of a heavy particle of mass $\Lambda \, \gg \, T_H$ is exponentially suppressed by a 
 Boltzmann  factor $\sim {\rm e}^{-{\Lambda \over T_H}}$.  Thus, our black hole with $N$ units of the  $Z_2^N$-charge, can start emitting $N$-species particles, only after its temperature drops to $T_H  \, \sim \, \Lambda$. At this point, the mass of the black hole is $M_{BH}^* \, \sim  \, {M_P^2 \over \Lambda}$.  Starting from this moment, the 
 black hole can start revealing back the stored charge, in form of the $N$-species particles. However, by conservation of energy, the  maximum number of particles that can be emitted by the black hole is 
 \begin{equation}
\label{nmax}
n_{max} \, \sim \, {M_P^2 \over \Lambda^2}.
\end{equation}
These states should carry the same $Z_2^N$-charge  as the original $N$-particles. Thus, $n_{max} \, = \, N$,  which proves the equation (\ref{PM}).  

In the other words, the key point of the proof, is that  the  amount of the maximal discrete charge that is stored in the initial black hole scales as $N$, but  the temperature  at which the black hole starts giving back this charge essentially {\it does not} scale with $N$. Hence the only way to avoid inconsistency 
is the scaling of the Planck mass$^2$  as  $\sim N$.

\subsection{ln$N$-corrections} 

  We should stress that the limit (\ref{PM}) must be understood  to leading order in large $N$, up to corrections that behave as ln$N$. To see this let us go through the black hole evaporation process 
 more carefully.   The rate of production of a particle of mass $\Lambda$ for a black hole 
 of the Hawking temperature $T_H \, \ll \, \Lambda$ is 
\begin{equation}
\label{rate}
\Gamma \, \sim \,   T_H {\rm e}^{- {\Lambda \over T_H}}.
\end{equation} 
 The total number of such particles from all the N-species produced during a time $t_*$ is 
\begin{equation}
\label{total}
n(t_*) \, \sim \, N \, \int_0^{t_*} \, T_H {\rm e}^{- {\Lambda \over T_H}}  \, dt \,.
\end{equation}
Noting that the cooling rate of the black hole (in the other species of mass $\ll T_H$)  is
\begin{equation}
\label{crate}
{dM_{BH} \over dt} \, \sim \, -  \, T^2_H,  
\end{equation} 
and using the relation between the black hole mass and the temperature $T_H \, \sim \, {M_P^2/M_{BH}}$,
we can re-express (\ref{total}) as an integral over the  temperature
\begin{equation}
\label{total}
n(t_*) \, \sim \, N\,  \int_{T_{in}}^{T_*} \, dT_H \, {M_P^2 \over T_H^3} \, {\rm e}^{- {\Lambda \over T_H}},  
\end{equation}
where $T_{in}$ is the initial temperature, which for the minimal black hole of interest is $T_{in} \, \sim  \, 
{M_P^2 \over N\Lambda}$.  Since we are proving the bound  (\ref{PM}), it is enough to show that the opposite
assumption leads us to the contradiction. So assume that  $M_P$ could violate the bound (\ref{PM}) so that  we could have
$M_P^2 \, \ll \, N \Lambda^2$.  Then 
$T_{in} \, \ll  \, \Lambda$, and the initial emission of $N$-species is exponentially suppressed. We now integrate (\ref{total}) to $T_*$ defined by the condition that $n(t_*) \lsim 1$. In the other words, we choose $T_*$ to be a temperature defined by the requirement that before reaching it the black hole  managed to emit only of order one number of  particles from the entire variety of the $N$-species.  

  Obviously, we have
$T_* \sim  \Lambda /{\rm ln}\left ({NM_P^2\over\Lambda^2}\right)$, or the corresponding mass $M^*_{BH}$ is 
\begin{equation}
\label{massstar}
M_{BH}^{*} \, \sim \, {M_P^2 \over \Lambda }{\rm ln}\left ({NM_P^2\over\Lambda^2}\right)\, ,
\end{equation} 
which corrects the bound (\ref{PM})  by a  factor   $\sim$ ln ${NM_P^2 \over \Lambda ^2}$, or equivalently by a factor  $\sim $ln$N^2$.

\section{Bound on $M_P$ for the Unstable $N$-Species}

 We shall now repeat our reasoning for the case in which $N$-species are not exactly stable and can decay into some lighter states, e.g., such as the light Standard Model particles.   Let the lifetime of the species be 
 $\tau_N$. Our proof of  the equation (\ref{PM}) then can still be applied as long as this lifetime
 is longer than the lifetime of the minimal black hole carrying the $N$-units of the $Z_2^N$ charge. 
 The lifetime of such a black hole is 
 \begin{equation}
\label{tbh}
\tau_{BH} \, \sim  \, {M_{BH}^3 \over M_P^4} \, = \, {N^3\Lambda^3 \over M_P^4}.
\end{equation}
Thus, as long as 
  \begin{equation}
\label{condition}
\tau_N \, \gsim \,  {N^3\Lambda^3 \over M_P^4}, 
\end{equation}
 our proof should hold, and we have a limit on $M_P$ given by (\ref{PM}). Substituting this in to the 
 (\ref{condition}), we get a bound on $\tau_N$ in terms of $N$ and $\Lambda$, for which our proof is still applicable
   \begin{equation}
\label{conditiontau}
\tau_N \, \gsim \,  {N \over \Lambda}.
\end{equation}
This expression is interesting because of the following reason.  First, observe that the lower bound of this expression is achieved if $N$-species decay into the lighter states through the operators suppressed by the powers of the scale $M \, \equiv  \,  \sqrt{N} \Lambda$, which, according to (\ref{PM}), saturates the lower bound on  the Planck mass.    Secondly, the same scale $M$ is independently
suggested by the perturbative arguments as the minimal scale suppressing the strength of the interactions of any representative of the  system of  $N$ inter-coupled species, because of the $N$-fold contribution into the  renormalization of their wave-functions.

\section{Black Hole Constraints on the Conserved Quantum Numbers}
  
 Our analysis has immediate implication for the conserved charges of arbitrary sort that are not characterized by any locally-observable  long-range (massless) fields. For example,  such are the charges conserved due to  the gauged discrete symmetries.  
 Our proof implies that any exactly conserved quantum number of the above sort (call it $Q$)  must be defined modulo (be periodic)  $N_{max}$, with  
 \begin{equation}
\label{Nbound}
N_{max} \, \lsim  \, \left ({M_P \over m}\right )^2, 
\end{equation}
where $m$ is the mass of a particle carrying one unit of the $Q$-charge.   We shall prove this bound
for the simplest case  when there is a single species carrying the charge $Q$, generalization to more complicated cases is straightforward. In the latter situation when there are multiple species carrying the 
different  amount of the  $Q$ charge, what counts is the mass to charge ratio. 
 
  We can again perform the similar thought experiment as for $N$-species, but instead of putting together  $N_{in}$ particles from different species, we shall form a black hole by putting together $N_{in}$ units of the $Q$-charge. Since each unit carries a mass $m$ the minimal mass of such a black hole is  $M_{BH} \, = \, N_{in}\, m$.   Since we wish to prove that $Q$ must have a limited periodicity
  of $N_{max}$  given by (\ref{Nbound}), it is enough to show that  the opposite assumption leads us into a contradiction. 
  
   Thus, we assume that the charge $Q$ has an unlimited  periodicity.  That is, can be arbitrarily large 
   in terms of the elementary charge. 
   Then,  by increasing $N_{in}$, we can store an arbitrarily  large amount of the $Q$ charge into the black hole.  But then again, the 
 black hole cannot return  back any significant amount of the  $Q$-charge up until it evaporates down to the mass $M_{BH}^* \, \sim  \,  {M_P^2 \over m}$, at which point its Hawking temperature becomes comparable to the mass of  an elementary  $Q$-quanta.  
  
  After this moment,  however, by conservation of energy, the black hole  can only give back $N_{max} \,  \sim \,  M_P^2/m^2$ units of the $Q$-charge, which contradicts to the initial condition that we could make initial $Q$-charge of the black hole 
 arbitrarily large, by increasing  $N_{in}$ unbounded.  The only way the story can be made consistent,  is if $Q$-charge is defined modulo $N_{max}$. In such a case irrespective of the 
 initial number of charged quanta,  the total charge of the black hole cannot exceed $N_{max}$, and 
 inconsistency is avoided. 
 
   One can come up with many situations for which the above proof  puts important constraints, and 
   we shall now briefly review some of them. 
   
   \subsection{Consistency Relation Between the Planck Mass and the Size of  a $Z_{N}$ Gauge Symmetry} 
   
    For example consider a single complex scalar field  $\Phi$ of mass $m$ transforming under a discrete gauge symmetry $Z_N$
  \begin{equation}
\label{zn}
\Phi \,  \rightarrow \, {\rm e}^{i{2\pi \over N}}\,  \Phi,  
\end{equation} 
Then, the maximum value of $N$ is  limited by (\ref{Nbound}).    
  
We can turn the argument around, and say that the presence of a  $Z_N$-symmetry at any scale 
$\Lambda$, implies the existence of the hierarchy between the scale $\Lambda$ and the Planck mass given by (\ref{PM}),  in which  $N$ has to be understood,  as the periodicity of the $Z_N$ group.

\subsection{Maximal Allowed $Z_N$-subgroups, of the Continuous Global  Symmetries}
   
Another important implication is for the continuous global  symmetries. Consider, for example,  a
single scalar field $\Phi$ of mass $m$ transforming under a global  $U(1)$ `baryon number' symmetry. 
If  $U(1)$ were exact, there would be a conserved baryon number $Q$, with unlimited periodicity.  
According to the above proof, however, this is impossible, and $Q$ can only be defined modulo 
$N_{max}$ given by  (\ref{Nbound}).  This means that any would be continuous global symmetry 
must be inevitably broken by the gravitational effects, at least down to the $Z_{N_{max}}$ subgroup.
That is, in the effective low energy theory, gravity should generate $U(1)$-violating operators
of the form
\begin{equation}
\label{break}
\Phi^N\, + \, \Phi^{*N}
\end{equation}
with $N \, \leq \, N_{max}$.

\subsection{Quantization of the Black Hole Massive Quantum Hair}

Finally, the bound (\ref{Nbound}) also implies the quantization of the black hole quantum hair under the massive 
integer spin gauge fields.  

 It was pointed out recently\cite{quantum}  that in the presence of spin-2 or higher integer-spin massive fields, the black holes may be endowed with the quantum hair under these fields. 
The field configuration is locally pure gauge,  and because of this it is clasically unobservable in full 
agreement with the standard no-hair arguments\cite{nohair, nohair1, nohair3}  and in particular the classical result  by Bekenstein \cite{nohair}.  However, the gauge field configuration has a  global topological structure, which makes it observable quantum-mechanically.  
In particular,  such a hair  can be detected at infinity by the Aharonov-Bohm \cite{AB} type 
experiment.  

 The essence of this quantum hair can be understood in the following way (see \cite{quantum} for the details).  The  massive high integer spin fields (e.g., spin-2) include a spin-1 "longitudinal' component  $\mathcal{A}_{\mu}$, which playes the role of a Goldstone-St\"ckelberg field for maintaining the gauge invariance in the presence of the mass.  Under the gauge transformation, $\mathcal{A}_{\mu}$ shifts as  
\begin{equation}
\label{gaugeA}
\mathcal{A}_{\mu} \, \rightarrow \, \mathcal{A}_{\mu} \,  - \, \xi_{\mu}, 
\end{equation}
where $\xi_{\mu}$ is an arbitrary regular vector. A  black hole with a quantum hair corresponds to the 
solution for which the spin-1 component  has a form of  Dirac's magnetic monopole \cite{quantum}
\begin{equation}
\label{dmonopole}
F_{ij} \, = \, \mu \, \epsilon_{ijk} {x^k \over r^3}, ~~~~~F_{0j} \, = \, 0,
\end{equation}  
where 
$F_{\mu\nu} \, = \, \partial_{\mu}\mathcal{A}_{\nu} \, - \, \partial_{\nu}  \mathcal{A}_{\mu}$, 
$x_j$ are space coordinates and $\mu$ is the quantum charge. However, unlike the would be Dirac's 
magnetic monopole,   this configuration is locally pure gauge,  since  $\mathcal{A}_{\mu}$ is eaxctly compensated by the other components of the massive 
high spin field, so that the full field is identically zero.  This is why it cannot be probed clasically by any local experiment.  However, it can be probed quantum mechanically provided there are boundary terms of the following form 
  \begin{equation}
\label{vectorandstring}
q\int dX^{\mu} \wedge dX^{\nu} \, F_{\mu\nu}. 
\end{equation}
This term describes a boundary coupling of the St\"uckelberg $\mathcal{A}_{\mu}$ to a  test string,  where   
$X^{\mu}$ are the string target space coordinates. $q$ is a constant. Since we are interested in large distance effects, the microscopic nature of the string is unimportant. Because (\ref{vectorandstring}) 
is a boundary term, it  is not affecting dynamics at the classical level,  but quantum mechanically 
can lead to an observable  effect in the presence of the quantum hair (\ref{dmonopole}). 
The string can detect the quatum hair of the black hole, by the Aharonov-Bohm experiment, in which the string loop lassoes the blach hole. 
The phase shift resulting from such an experiment is  
\begin{equation}
\label{shift1}
{\rm phase~shift}\, = \, 4\pi \mu q,
\end{equation}  
and is observable as long as  $\mu q \, \neq \, n/2$. 
Thus, black holes can have an additional locally-unobservable quantum
charge  $Q \equiv 2 \mu q$.
 We shall now show that our black hole arguments impose the  following quantization condition on $Q$.

  Let the mass of a black hole (or a particle),  that carries an elementary unit of the quantum charge $Q$, be $m$.  Naively,  this  quantum charge $Q$ can be arbitrarily small.  Then,  
because  $Q$ causes no locally-observable long-range field, there is no obstruction in putting together  an arbitrary number $N$ of the unit charges and producing a black hole 
that would store $NQ$ quantum charge, which should be given back after evaporation.
If $Q$ could be arbitrarily small, then $N$ could be arbitrarily large. 
However,  according to our proof this is impossible.  Which implies inevitable periodicity 
of  $Q$ set by  (\ref{Nbound}).  
 Or in the other words quantization of $Q$ in the units of  $1 \over N_{max}$.

\section{Implications for the Hierarchy Problem and LHC} 

 From the above analysis, there emerge the  two possible solutions to the hierarchy problem. 
 
 The first approach  is to guarantee weakness of gravity by postulating a discrete symmetry, with a huge periodicity, around the TeV scale. 
 
For instance,  we may postulate that beyond the Standard Model there is an additional complex scalar
$\Phi$ with a $\sim $TeV mass, that transforms under some  $Z_N$-symmetry  with $N \, = \, 10^{32}$.  Then, this fact  would automatically imply the needed weakness of gravity. 

 Another approach is to simply postulate that there are additional $10^{32}$ species on top of the standard model, or even $10^{32}$ replicas of the Standard Model, all related by a  permutation  symmetry.  
 
  Although seemingly different, the connection between these two approaches may be stronger than what one would naively guess.   The reason is, that in many instances generation of large discrete
 symmetry groups implies the existence of  many species in the underlying theory, or else we may violate some conditions of our theorem, and weakness of gravity will no longer be guaranteed.    
 To illustrate this point, let us consider an example in which  one wants to generate an effective
  $Z_N$-symmetry, with $N \sim 10^{32}$,  from a renormalizable theory with order one couplings.  We can achieve this by 
 introducing a set of $n$ scalar fields $\Phi_k, ~~k \, = \, 1,2,...n$, with the following sequence of couplings 
  \begin{equation}
\label{sec}
\Phi_1^3 \, + \, \Phi_1^* \Phi_2^3 \, + \, \Phi_2^* \Phi_3^3 \, + \, ... \,  \Phi_k^*\Phi_{k\, +\,1}^3 \, +  ...
\, + \, \Phi_{n\, -\,1}^*\Phi_n^3 \, ,
\end{equation}
where under  $Z_{3^{n}}$,
\begin{equation}
\label{zncharge}
\Phi_k \, \rightarrow \, {\rm e}^{i{\pi \over 3^{k}}} \Phi_k \, .
\end{equation}
Integrating out all $n-1$ fields, the effective coupling for $\Phi_n$ will be 
\begin{equation}
\label{effectiven}
\Phi_n^{3^n} \, + \, h.c., 
\end{equation}
which exhibits, $Z_{3^{n}}$ -invariance. The above construction  leaves an impression that we have managed to generate a discrete  $Z_N$-symmetry with exponentially  large  $N = 3^n$,  by  introducing just  $n$ fields. 
This is true, however there is the following  caveat as far as the solution of the hierarchy problem is conserved.  If we don't want to have any input hierarchy  of scales,  the masses of all the 
integrated out fields must not be much above $\Lambda$.  However, in the latter case, the condition of our theorem is violated, since fields with $k\,  < \, n$ will inevitably have much larger $Z_{3^n}$-charge-to-mass ratio 
than $\Phi_n$.  A black hole prepared  with an arbitrarily large number of  $\Phi_n$ fields can get rid of its $Z_{3^n}$-charge by emitting just  $\sim \, n$ number of $\Phi_k$-quanta with needed values of $k$, 
without implying large $M_P/\Lambda$  hierarchy. 
 
  The lesson from the above example is, that generation of large discrete symmetries without violating any of the conditions of the bound (\ref{PM}),  may require either an unnaturally small coupling, or the large number of fields.   For example,  $N$ fields with exact permutation symmetry, would do such a job.

 One way or the other, we are lead to the conclusion that the large $N$ solution of the hierarchy 
 problem (whether  $N$ refers to the number of species or to the periodicity of the $Z_N$ symmetry)
 implies the existence of a sector beyond the Standard Model with enormous number of 
 quantum fields.  This fact is the key for the testability  of the above proposal. 
  We shall now argue, that under very reasonable  assumptions this sector will be probed by 
  LHC. 
  
  Since the $N$-sector states are singlets under the standard model gauge group, their couplings 
  to the SM sector must be through the effective non-renormalizable  interactions. The exceptions are the spin-1 and spin-0 fields. 
  The former could couple, for instance, to baryon or lepton number currents, and  the latter  
  could have a renormalizable couplings with the Standard Model Higgs  doublet.  Also, gauge-neutral 
  neutrinos from the different species could also mix.   
  
  Renormalizable or not,  the strength of the $N$-species interaction  with the Standard Model
  fields  must be suppressed  
by the powers of the Planck mass, provided most of them  couple to our sector and to each other.   In the opposite case one has to explain why the standard model fields themselves are not weakly coupled, since the $N$-sector 
loops would lead to large wave function renormalization of the Standard Model  fields, just like  they do for the graviton. 

  Hence,  we expect the $N$-sector species to couple with the standard model through the 
  effective operators of the form
  \begin{equation}
\label{ncoupling}
\sum_i {\Phi_i \over M_i}  T_{\mu}^{\mu}  \, + \,  \sum_i {\Phi^{(i)}_{\mu\nu}  \over \bar{M}_i} T^{\mu\nu} 
\, + ... \, . 
\end{equation}
Here we gave the  examples of spin-0 and spin-2 species respectively, and $T_{\mu\nu}$ is the  energy
momentum  
tensor of the Standard Model fields.   According to our arguments, scales $M_i, \bar{M}_i$ must scale 
as $\sim N$, and thus, be of the order of $M_P$.   The $N$-species then will be produced in 
the collision of SM particles, and in particular at LHC. Although the production rate of any particular species is gravitationally suppressed, their number overcompensates.  The production rate 
of all the species above their mass, in some collision process at typical energy $E$,  is 
 \begin{equation}
\label{raten}
\Gamma_{total} \, \sim \, {E^3 \over M_P^2} N \, \sim  \ {E^3 \over \Lambda^2},  
\end{equation}
where in the last expression we  have used the bound (\ref{PM}).  This expression, is the part of the general rule, which states the following.  By allowing the natural strength (which is also a maximal allowed strength) coupling 
to the  $N$-species that maintain the  hierarchy between the 
Planck mass and any scale $\Lambda \, \ll \,  M_P$, their production rate becomes important at the scale $\Lambda$.  Since in our case  $\Lambda \sim $TeV, the relevance for the LHC automatically 
follows. 

  We wish to point out some resemblance between (\ref{raten}) and the production rate of 
  the Kaluza-Klein  gravitons in Large Extra Dimensional model \cite{ADD}.  In the light of our discussion, this is not at al surprising, since   Large Extra Dimensions solution of the hierarchy problem, can be regarded as  a particular example of  the large $N$-solution, due to the fact that it also includes $\sim 10^{32}$ Kaluza-Klein gravitons  of mass $\sim$ TeV.  Thus, even if one knew nothing about the high-dimensional Gauss law and the relation between the fundamental and four-dimensional Planck masses, 
 the solution of the hierarchy problem, in the light of the presented black hole  arguments, becomes obvious the moment  one realizes that there are $10^{32}$  Kaluza-Klein species around the TeV scale.

  Finally,  some open questions are in order.   The natural one to ask is,  what  is the connection between  the bound (\ref{PM})  Bekenstein-Hawking entropy. 
  
  It also would be interesting to explore the connection between the work \cite{nima} and the arguments presented here.   
  
  $~~~~~~$

  {\bf Acknowledgments}

We thank Gregory~Gabadadze, Matt~Kleban,  Michele~Redi and Oriol Pujolas  for very useful discussions and comments. 
The work  is supported in part  by David and Lucile  Packard Foundation Fellowship for  Science and Engineering, and by NSF grant PHY-0245068.


\end{document}